\documentclass[showpacs,footnoteinbib]{revtex4}
\usepackage{amsmath,amssymb}
\usepackage{color}
\usepackage{graphicx}

\begin{document}

\title{Single electron quantum tomography in quantum Hall edge channels}

\author{Ch. Grenier$^{1}$}
\author{R. Herv{\'e}$^{1}$}
\author{E. Bocquillon$^{2}$}
\author{F.D. Parmentier$^{2}$}
\author{B. Pla\c cais$^{2}$}
\author{J.-M. Berroir$^{2}$}
\author{G.~F\`eve$^{2}$}
\author{P.~Degiovanni$^{1}$}
\email{Pascal.Degiovanni@ens-lyon.fr}
\
\affiliation{(1) Universit\'e de Lyon, F\'ed\'eration de Physique Andr\'e Marie Amp\`ere,
CNRS - Laboratoire de Physique de l'Ecole Normale Sup\'erieure de Lyon,
46 All\'ee d'Italie, 69364 Lyon Cedex 07,
France}

\affiliation{(2) Laboratoire Pierre Aigrain, D{\'e}partement de Physique
de l'Ecole Normale Sup\'erieure, 24 rue Lhomond, 75231 Paris Cedex
05, France}

\begin{abstract}
We propose a quantum tomography protocol to measure single electron coherence in quantum Hall edge channels and therefore
access for the first time the wave function of single electron excitations propagating in ballistic quantum conductors.
Its implementation would open the way to quantitative 
studies of single electron decoherence  and would provide a quantitative tool for analyzing single to few electron sources. We show
how this protocol could be implemented using ultrahigh sensitivity noise measurement schemes.
\end{abstract}


\maketitle

\section{Introduction}
\label{sec:introduction}

Electron quantum optics aims at the controlled manipulation and measurement of the quantum state of a single to few 
electrons propagating in solid state in a comparable way to the recent achievements with microwave photons \cite{Deleglise:2008-1,Hofheinz:2009-1} 
or light \cite{Ourjoumtsev:2007-1}. In particular, it requires the use of ballistic quantum conductors were single electrons 
can propagate along one dimensional modes. Using continuous sources, Mach Zehnder interferometers have 
been realized in integer quantum Hall edge channels demonstrating single electron  \cite{Ji:2003-1,Roulleau:2008-2,Neder:2007-4}  as well as two electron
\cite{Neder:2007-2} quantum coherence following a proposal by Samuelsson, Sukhorukov and B\"{u}ttiker \cite{Samuelsson:2004-1}. 
Recently an on-demand single electron and single hole source based on the mesoscopic capacitor has been demonstrated  \cite{Feve:2007-1,Mahe:2010-1}.
A two terminal single electron and hole source based on a dynamical quantum dot \cite{Leicht:2011-1} operating at GHz frequencies has also been demonstrated
in quantum Hall edge channel, as well as a similar electron pump in a 2D electron gas (2DEG) in zero magnetic field \cite{Blumenthal:2007-1}. Single electron excitations 
can also be generated at GHz repetition rates using surface acoustic waves \cite{Talyanskii:1997-1,Ahlers:2006-1} and detected with high efficiency 
after propagation within a 1D non chiral channel \cite{Hermelin:2011-1}. All these developments in quantum transport and single electron electronics have risen the hope for electron quantum 
optics experiments involving single electron excitations \cite{Olkhovskaya:2008-1,Splettstoesser:2009-1}.

In this context, it is important to understand precisely the similitudes and differences between electron quantum optics and 
photonic quantum optics. The Fermi statistics of electrons is expected to bring in new features. First of all, the 
ground state of a metallic conductor is a Fermi sea characterized by its chemical potential.  
Fermi sea vacua have radically different properties from the photon vacuum since,
due to Fermi statistics, entanglement can be generated by sources at equilibrium even in the absence of
interactions \cite{Beenakker:2006-1}. Besides quantum statistics, 
Coulomb interactions lead to decoherence of electronic excitations whose consequences
in the context of Mach-Zehnder interferometers have been extensively discussed in the 
recent years \cite{Chung:2005-1,Chalker:2007-1,Neder:2007-3,Neuenhahn:2008-1,Levkivskyi:2008-1,Roulleau:2008-1,Kovrizhin:2009-1}. 

However, with the advent of on demand single electron sources, 
the problem of electronic decoherence has to be reconsidered. A crucial question is to understand the 
deviation from the non-interacting picture for single electron excitations emitted by these new sources
due to electron-electron interactions and to decoherence induced by the electromagnetic environment. 
As suggested by recent experimental studies of electron relaxation in
quantum Hall edge channels \cite{Altimiras:2010-1,LeSueur:2010-1} these effects seriously question 
the quantum optics paradigm of  electronic quasiparticles in quantum Hall edge channels. Although this problem 
has been investigated on the theoretical side \cite{Degio:2009-1}, it is important to develop new experimental tools that allow to tackle these 
issues in a direct and accessible way.

\medskip

For this reason, we propose a quantum tomography protocol for single electron excitations in quantum Hall edge channels
in the spirit of homodyne tomography in quantum optics \cite{Smithey:1993-1,Lvovsky:2009-1}. 
Despite the recent experimental achievements in electron quantum optics, 
the quantum state of a single electron excitation has never been imaged and our proposal is designed to fill that gap. 
Performing such a single electron quantum tomography 
would then open the way to experimental studies of single electron decoherence in 
nanostructures and consequently to quantitative tests of theoretical approaches to this basic problem \cite{Degio:2009-1}. Our 
single electron quantum tomography protocol 
would also  lead to a new characterization of the quantum coherence properties of single to 
few electron sources \cite{Feve:2007-1,Leicht:2011-1,Keeling:2006-1}.

To support its feasability, we discuss predictions for the experimental signals produced by a realistic source of energy resolved single 
electron excitations emitted by the mesoscopic capacitor \cite{Feve:2007-1}. Our study
shows that this proposal could be implemented using 
recently developed ultrahigh sensitivity noise measurement schemes \cite{Parmentier:2010-1}. We also discuss 
quality assessment for the coherence of single electron sources in terms of quantum information concepts
such as linear entropy and fidelity with respect to trial single electron wavefunctions. We show how these quantities can
be accessed through a full quantum tomography of single electron excitations.

\medskip

This paper is structured as follows: the notion of single electron coherence is recalled and briefly discussed in section \ref{sec:electron-coherence}.
Our proposal for a single electron quantum tomography protocol is then described in section \ref{sec:tomography}. Finally, predictions and quality assessment
for the on demand single electron source are presented in section \ref{sec:electron-source}.

\section{Single electron coherence}
\label{sec:electron-coherence}

For a many body system, the quantum coherence properties at the single particle level are encoded within the 
space and time dependence of the 
two point Green's function, called the single electron coherence and analogous of field correlations introduced by 
Glauber for photons \cite{Glauber:1963-1}:
\begin{equation}
\mathcal{G}^{(e)}(x,t;x',t') = \langle \psi^\dagger(x',t')\psi(x,t)\rangle
\end{equation}
where operators $\psi(x)$ and $\psi^\dagger(x)$ destroy and create a single electron at position $x$. In the same way, 
the single hole coherence is defined as 
\begin{equation}
\mathcal{G}^{(h)}(x,t;x',t') = \langle \psi(x',t')\psi^\dagger(x,t)\rangle\,.
\end{equation}
In this paper, we consider ballistic conductors formed by a single quantum edge Hall channel. Thus, electron propagation
within the edge channel is chiral at the Fermi velocity $v_F$ so that  the single electron coherence obeys this property. Since measurements
are usually made at a given position, we will focus on the time dependence 
at a given position which by chirality also characterizes spatial coherence 
properties. 

In full generality, the single electron coherence can be decomposed as the sum an equilibrium
contribution $\mathcal{G}^{(e)}_{\mu}$ due to the Fermi sea $|F_{\mu}\rangle$ at electrochemical potential  $\mu$ 
and an extra contribution $\Delta\mathcal{G}^{(e)}(t,t')$ representing single particle coherence of excitations emitted by sources 
within the conductor:
\begin{equation}
\label{eq:4.0}
\mathcal{G}^{(e)}(t,t')=\mathcal{G}_{\mu}^{(e)}(t-t')+\Delta\mathcal{G}^{(e)}(t,t')\,.
\end{equation}
Note that any stationary single particle coherence such as $\mathcal{G}_{\mu}^{(e)}$ only depends on $t-t'$ and not
on $\bar{t}=(t+t')/2$. Since we are interested in single electron sources that produce a 
non stationary single electron coherence, the $\bar{t}$ dependence
of $\Delta\mathcal{G}^{(e)}(t,t')$ must be retained. Note that $-ev_{F}\Delta\mathcal{G}^{(e)}(t,t)$ is the 
average excess current measured at position $x$ and time $t$.

\medskip

As an example, let us consider an ideal one shot single electron 
source that would inject an electronic excitation in wavepacket $\varphi_{e}$ above the Fermi sea, {\it i.e.}
such that in the frequency domain $\varphi_{e}(\omega)=0$ for $\omega\leq 0$. This ideal source would 
generate the many-body state 
\begin{equation}
\label{eq:sec:single-electron-state}
\psi^\dagger[\varphi_{e}]\,|F_{\mu}\rangle=
\int \varphi_{e}(x)\psi^\dagger(x)|F_{\mu}\rangle\,dx\,.
\end{equation}
Then, using Wick's theorem, the single electron coherence at $x=0$ due to the single excitation can be readily evaluated:
\begin{equation}
\label{eq:sec:ideal-coherence}
\Delta\mathcal{G}^{(e)}_{\psi^\dagger[\varphi_{e}]\,|F_{\mu}\rangle}(t,t')=\varphi_{e}(-v_{F}t)\,\varphi_{e}^*(-v_{F}t')\,.
\end{equation}

\begin{figure}
\begin{center}
\includegraphics[width=12cm]{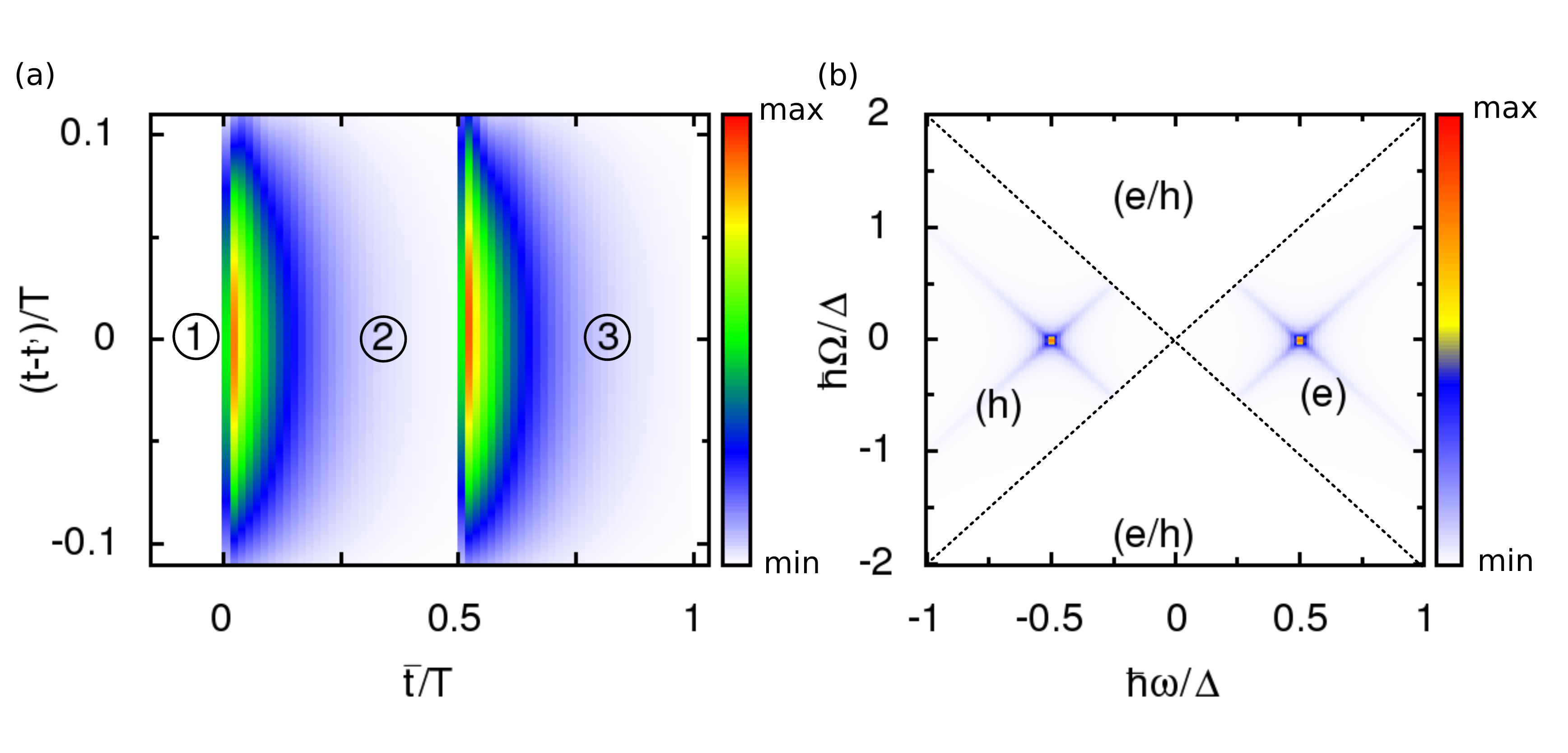}
\end{center}
\caption{\label{fig:coherence} 
Expected single electron coherence from the single electron source \cite{Feve:2007-1}
at the electron hole symmetric point: $\hbar\omega_{e}=-\hbar\omega_{h}=\Delta/2$:
(a)  $|v_{F}\Delta\mathcal{G}^{(e)}(t,t')|$ from the discrete level coupled to a continuum 
approach ($h\gamma_{e}/\Delta\simeq 0.21$ and $T\gamma_{e}\simeq 6.85$). 
(b) Modulus of $\Delta\mathcal{G}^{(e)}$ in Fourier space: 
$\omega$ is conjugated to $t-t'$ and $\Omega$ to $\bar{t}=(t+t')/2$. 
Quadrants $(e)$ (resp. $(h)$)
represent electron (resp. hole) coherence whereas the $(e/h)$ quadrants encode electron/hole coherence. Energy resolved 
single electron (respectively hole) excitations can be seen in the $(e)$ (respectively $(h)$) quadrants.
The horizontal line $\Omega=0$ gives the average excess electron occupation number due to the source.}
\end{figure}

This example shows that an experimental determination of $\Delta\mathcal{G}^{(e)}$
provides a direct visualization of wavefunctions of coherent single electron excitations.
However, in any real device, many particle correlations due to the Pauli exclusion principle \cite{Keeling:2008-1} and Coulomb interactions 
may lead to relaxation and decoherence of single electron or hole excitations \cite{Degio:2009-1}. Then, because of decoherence,
$\Delta\mathcal{G}^{(e)}(t,t')$ is not of the simple form given by \ref{eq:sec:ideal-coherence}.
Nevertheless, its behavior in $|t-t'|$ still describes the temporal coherence of the electrons at the position of measurement
and thus provides information on their coherence time. Accessing the coherence properties of energy resolved
single electron excitations is crucial for probing the chiral Fermi liquid paradigm in quantum Hall edge channels
in the spirit of Landau's original discussion of the quasiparticle concept \cite{AbrikosovGorkov}. 

\medskip

In practice, since sub-nanosecond detection of a single electron cannot be achieved in the present status of
technology, it is more convenient to access the coherence of single particle excitations in the frequency domain:
\begin{equation}
\label{eq:4}
\Delta\mathcal{G}^{(e)}(t,t')=\int \Delta\mathcal{G}^{(e)}(\omega_{+},\omega_{-})\,e^{i(\omega_-t'-\omega_{+}t)}\,\frac{d\omega_{+}\,d\omega_{-}}{(2\pi)^2}\,.
\end{equation}
where $\omega_{+}$ and $\omega_{-}$ are respectively conjugated to $t$ and $t'$. In the frequency domain, the stationary part is
encoded within the diagonal $\omega_{+}=\omega_{-}$ whereas the non stationarity of single electron coherence is encoded
in its $\Omega=\omega_{+}-\omega_{-}$ dependence. Let us remember that electron distribution function measurements \cite{Altimiras:2010-1} 
only give access to the stationary part of the single electron coherence (diagonal $\omega_{+}=\omega_{-}$ or equivalently $\Omega=0$) but miss to capture its $\bar{t}=(t+t')/2$ dependence
encoded in the off diagonal terms of the single electron coherence in frequency space: $\mathcal{G}^{(e)}(\omega_{+},\omega_{-})$ for $\omega_{+}\neq \omega_{-}$.

\medskip

\begin{figure}
\begin{center}
\includegraphics[height=6cm]{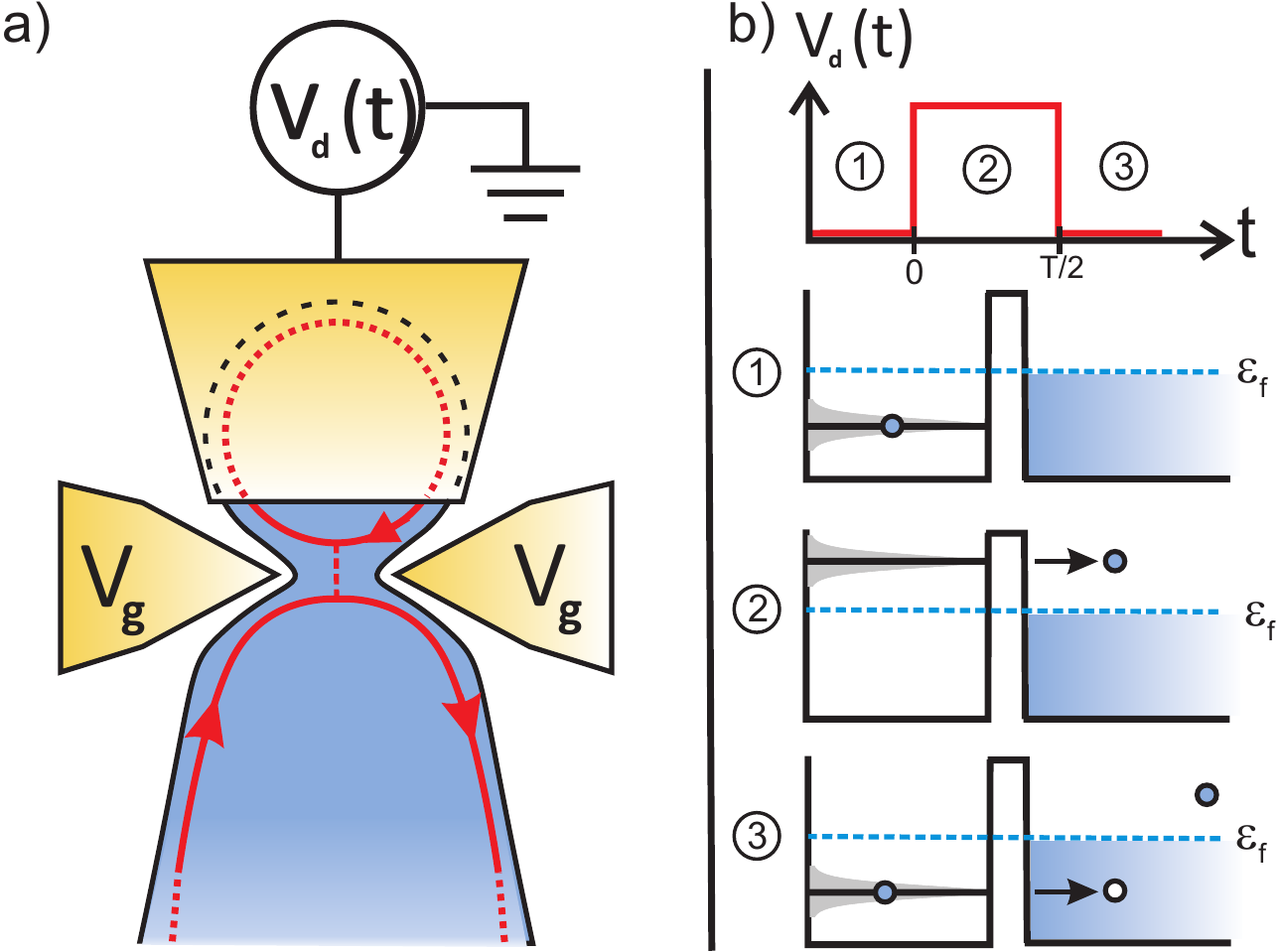}
\end{center}
\caption{\label{fig:source} (a) A micron-sized 
dot with level spacing $\Delta$ and a tunable quantum point contact (QPC) is capacitively coupled to a top metallic gate. (b) Applying a square
voltage $V_{d}(t)$ to the top gate brings a populated dot level above the Fermi level, thus emitting
a single electron at energy $\hbar\omega_e=+\Delta/2$, followed by a single hole 
when the level is brought back at energy $\omega_h=-\Delta/2$ below the Fermi energy. 
The widths of these excitations $\gamma_{e,h}$ are given by the
corresponding escape rates from the dot.}
\end{figure}

Figure \ref{fig:coherence} presents density plots of single electron coherence that would 
be emitted by an ideal on demand single electron source \cite{Feve:2007-1} based on the mesoscopic capacitor depicted on figure \ref{fig:source}a.
Ideally, such a source should emit a single electron followed by
a single hole excitation: at $t=0$, the highest occupied energy level of a quantum dot  (see figure \ref{fig:source}b)
is moved at energy $\hbar\omega_{e}>0$ above the Fermi level (taken for simplicity at zero) and releases a single electron in the
continuum of available single particle states $\varphi_{\omega}(x)=e^{i\omega x/v_F}$ ($\omega>0$). The resulting
single particle wave function is obtained as a truncated Lorentzian in the frequency domain:
\begin{equation}
\label{eq:lorentzian}
\widetilde{\varphi}_{e}(\omega)= \frac{\mathcal{N}_e\,\theta(\omega)}{\omega-\omega_{e}-i\gamma_{e}/2}
\end{equation}
where $\mathcal{N}_{e}$ ensures normalization and $\gamma_{e}$ denotes the electron 
escape rate from the quantum dot. Hole emission starts at $t=T/2$ when electron emission is completed
($\gamma_{e}T\gg 1$) and is described in a similar way with the release of a single hole truncated Lorentzian wavepacket 
at energy $\hbar\omega_{h}<0$ in the
continuum of available hole states below the Fermi level. Ideally, the source is expected to release a single electron and a single hole and therefore to generate the
state $\psi^\dagger[\varphi_{e}]\psi[\varphi_{h}]\,|F_{\mu}\rangle$.
Figure \ref{fig:coherence}a presents a density plot for $|v_{F}\Delta\mathcal{G}^{(e)}(t,t')|$ for such a state as a function
of $\bar{t}=(t+t')/2$ and $t-t'$: the electron and hole wavepackets emitted during each half period are clearly seen. The $\bar{t}$ dependence for $t=t'$ is
the exponential decay of the average electrical current as observed and characterized experimentally \cite{Mahe:2008-1} 
and the decay of $|v_{F}\Delta\mathcal{G}^{(e)}(t,t')|$ along $|t-t'|$ reflects the truncation of the Lorentzian. 

Figure \ref{fig:coherence}b then
presents a density plot of $|v_{F}\Delta\mathcal{G}^{(e)}(\omega_{+},\omega_{-})|$ in function of $\omega=(\omega_{+}+\omega_{-})/2$ and 
$\Omega=\omega_+-\omega_{-}$.
Note that the quadrant (e) on figure \ref{fig:coherence}b, defined by both $\omega_{+}$ and $\omega_{-}$ positive, corresponds to single particle 
states with energy above the Fermi energy (electron states). Similarly, the quadrant (h) with both $\omega_{+}$ and $\omega_{-}$ negative corresponds to hole states.
Figure \ref{fig:coherence}b clearly exhibits energy resolved electron and hole excitations.

The off diagonal quadrants (e/h) on figure \ref{fig:coherence}b are defined by $\omega_{+}\omega_{-}<0$ and correspond to 
electron/hole coherence. 
Such an electron/hole coherence appears in superpositions of states with different electron/hole pair numbers
such as, for example, $\alpha |F_{\mu}\rangle+\beta \psi^\dagger[\varphi_{e}]\psi[\varphi_{h}]|F_{\mu}\rangle $. In such a state, the single
electron coherence contains interference contributions of the form 
$\langle F_{\mu}|\psi^\dagger(x,t')\psi(x,t)\psi[\varphi_{h}]\psi^\dagger[\varphi_{e}]|F_{\mu}\rangle$
and $\langle F_{\mu}|\psi[\varphi_{e}]\psi^\dagger[\varphi_{h}]\psi^\dagger(x,t')\psi(x,t)|F_{\mu}\rangle$.
Using Wick's theorem and assuming as before that $\varphi_{e/h}$ are respectively pure electron and hole wavefunctions, 
we obtain for example that $\langle F_{\mu}|\psi^\dagger(x,t')\psi(x,t)\psi[\varphi_{h}]\psi^\dagger[\varphi_{e}]|F_{\mu}\rangle=
-\varphi_{h}(x-v_{F}t')^*\varphi_{e}(x-v_{F}t)$. This shows that these interference contributions live in the (e/h) quadrants of the frequency domain. 
Let us point out again that an ideal single electron and hole source should not exhibit electron/hole coherences.

\medskip

\section{Single electron quantum tomography}
\label{sec:tomography}

So far, in mesoscopic conductors,
a quantum tomography protocol has only been proposed for orbital states \cite{Samuelsson:2006-1},
but not for the reconstruction of temporal or spatial single particle coherence. To achieve this goal, 
we propose to use an Hanbury~Brown \& Twiss (HBT) setup \cite{Hanbury:1956-2,Liu:1998-1,Henny:1999-1,Oliver:1999-1} 
depicted on figure \ref{fig:HBT}a. Our proposal is based on an analogy with optical tomography \cite{Smithey:1993-1,Lvovsky:2009-1} and its simple design, also proposed
to realize a Hong-Ou-Mandel experiment \cite{Hong:1987-1} with single  electrons \cite{Olkhovskaya:2008-1}, makes it a potentially general 
tool for electronic quantum coherence measurement.

\subsection{The Hanbury Brown and Twiss effect}

\begin{figure}
\begin{center}
\includegraphics[width=12cm]{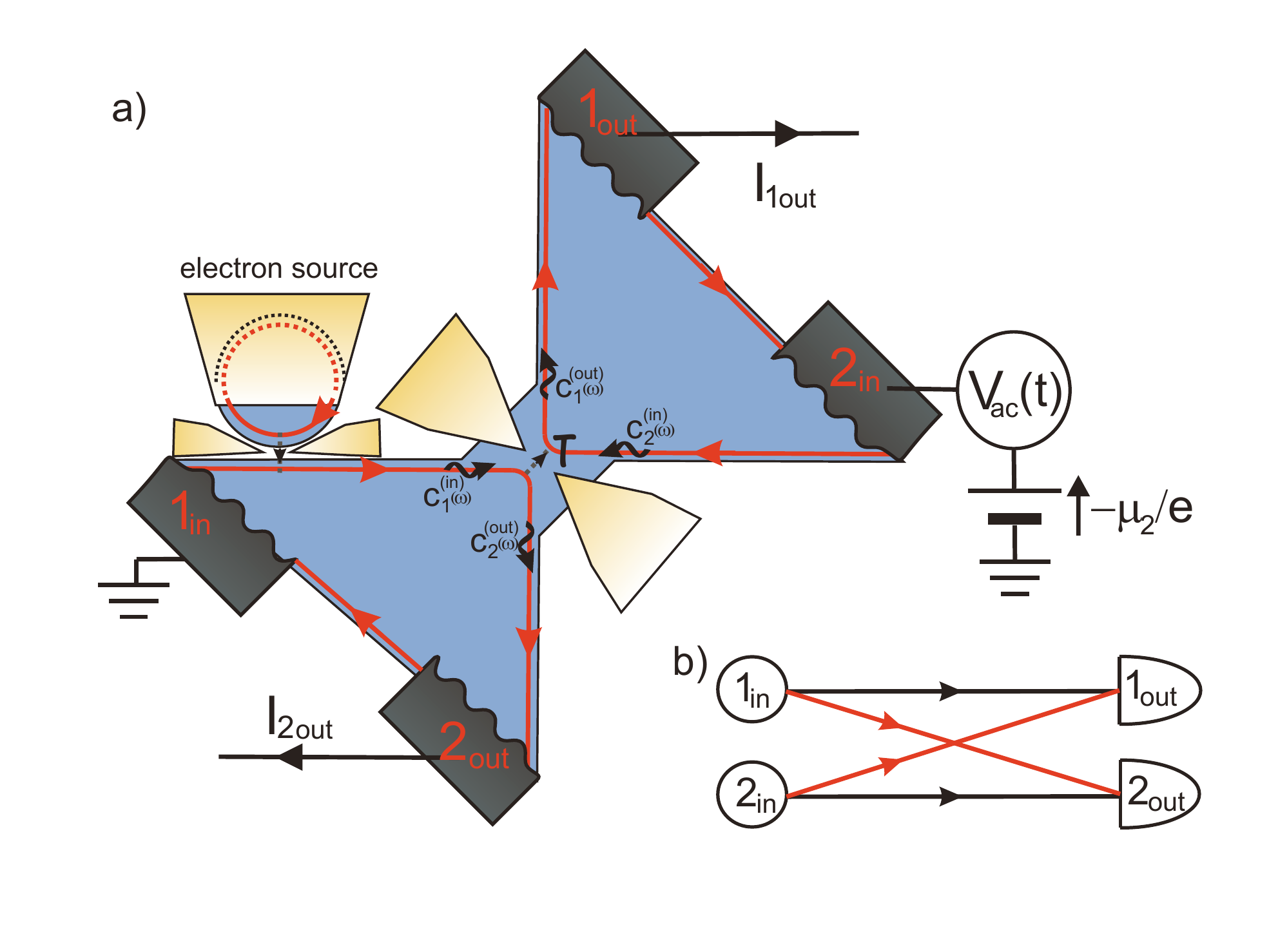}
\end{center}
\caption{\label{fig:HBT} (Color online) (a) The beam splitter is a quantum point contact
of transparency $\mathcal{T}$. The source is placed 
on incoming channel $1$ whereas channel $2$ at chemical potential $\mu_{2}$ can be driven by  
$V_{\mathrm{ac}}(t)=V\,\cos{(2\pi nft+\phi)}$ where $f=1/T$ is the driving frequency of the source. Outcoming
current correlations are measured at low frequency. (b) Direct and exchange two-particles paths responsible for the HBT effect.}
\end{figure}

The HBT effect arises from two particle 
interferences between direct and exchange paths depicted schematically on figure \ref{fig:HBT}b.
These interferences lead to the bunching of indistinguishible bosons and the antibunching of indistinguishible fermions expected from their quantum statistics.
As a consequence, when indistinguishible particles from two independent sources collide on a beam splitter,
the outcoming particle current fluctuations and correlations encode information on the single particle contents in the two incoming
beams.
Since its discovery \cite{Hanbury:1956-1,Hanbury:1956-2} the HBT effect has also been observed for electrons in a 2DEG issued by two different 
reservoirs at equilibrium \cite{Henny:1999-1,Oliver:1999-1,Liu:1998-1}. Here we discuss 
how the HBT effect manifests itself in outcoming current correlations in the HBT setup
depicted on figure \ref{fig:HBT}a.

\medskip

In this HBT setup, the quantum point contact (QPC) acts as a perfect electron beam splitter with energy independent
reflexion and transmission probabilities $\mathcal{R}$ and $\mathcal{T}$ ($\mathcal{R}+\mathcal{T}=1$). 
Let us introduce the incoming and outcoming electron modes within each channel ($\alpha\in\{1,2\}$) as depicted on 
figure \ref{fig:HBT}a:
\begin{eqnarray}
\label{eq:HBT:incoming}
\psi_{\alpha}^{(\mathrm{in})}(t) & = & \int c_{\alpha}^{(\mathrm{in})}(\omega)\,e^{-i\omega t}\frac{d\omega}{\sqrt{2\pi v_F}}\\
\psi_{\alpha}^{(\mathrm{out})}(t) & = & \int c_{\alpha}^{(\mathrm{out})}(\omega)\,e^{-i\omega t}\frac{d\omega}{\sqrt{2\pi v_F}}
\label{eq:HBT:outcoming}
\end{eqnarray}
where $\psi_{\alpha}^{(\mathrm{in})}(t)$ denotes the electron field in channel $\alpha$ right before the quantum point contact whereas the
$\psi_{\alpha}^{(\mathrm{out})}(t)$ are taken right after the QPC. The outcoming
electron modes are then related to the incoming ones by the QPC scattering matrix which we take of the form:
\begin{equation}
\label{eq:tomography:scattering}
\left(\begin{array}{c}
c^{(\mathrm{out})}_{1}(\omega)\\
c^{(\mathrm{out})}_{2}(\omega)
\end{array}
\right)=\left(
\begin{array}{cc}
\sqrt{\mathcal{T}} & i\sqrt{\mathcal{R}}\\
i\sqrt{\mathcal{R}} & \sqrt{\mathcal{T}}
\end{array}\right)
\,
\left(\begin{array}{c}
c^{(\mathrm{in})}_{1}(\omega)\\
c^{(\mathrm{in})}_{2}(\omega)
\end{array}\right)\,.
\end{equation}
Using this scattering matrix, the outcoming current operators can be expressed in terms of the incoming fermion fields. Therefore the outcoming current correlations 
defined as 
$S_{\alpha,\beta}^{\mathrm{out}}(t,t')=\langle i^{\mathrm{out}}_{\alpha}(t)\,i^{\mathrm{out}}_{\beta}(t')\rangle
-\langle i^{\mathrm{out}}_{\alpha}(t)\rangle\langle i^{\mathrm{out}}_{\beta}(t')\rangle$
($(\alpha,\beta)\in\{1,2\}$) can be computed in terms of incoming current and electronic correlations:
\begin{eqnarray}
\label{eq:QPC:S11}
S_{11}^{\mathrm{out}}(t,t') & = & \mathcal{R}^2S_{11}^{\mathrm{in}}(t,t')+\mathcal{T}^2S_{22}^{\mathrm{in}}(t,t') +\mathcal{RT}\,\mathcal{Q}(t,t')\\
\label{eq:QPC:S22}
S_{22}^{\mathrm{out}}(t,t') & = & \mathcal{T}^2S_{11}^{\mathrm{in}}(t,t')+\mathcal{R}^2S_{22}^{\mathrm{in}}(t,t') +\mathcal{RT}\,\mathcal{Q}(t,t')\\
\label{eq:QPC:S12}
S_{12}^{\mathrm{out}}(t,t')&  = & S_{21}^{\mathrm{out}}(t,t') = \mathcal{RT}\left(S_{11}^{\mathrm{in}}(t,t')+S_{22}^{\mathrm{in}}(t,t')-\mathcal{Q}(t,t')\right)\,
\end{eqnarray} 
where $S_{11}^{\mathrm{in}}(t,t')$ and $S_{22}^{\mathrm{in}}(t,t')$ denote the incoming current noises 
and $\mathcal{Q}(t,t')$ is the HBT contribution to outcoming correlations. Encoding two particle interferences, 
it involves incoming single electron and hole coherences at different times, right before the QPC:
\begin{equation}
\label{eq:Q}
\mathcal{Q}(t,t')=(ev_F)^2\left(
\mathcal{G}_{1}^{(e)}(t',t)\mathcal{G}_{2}^{(h)}(t',t)+
\mathcal{G}_{2}^{(e)}(t',t)\mathcal{G}_{1}^{(h)}(t',t)\right)\,.
\end{equation}
Equations \eqref{eq:QPC:S11}, \eqref{eq:QPC:S22}, \eqref{eq:QPC:S12}
and \eqref{eq:Q} suggest that putting a suitable source on channel 2 of the HBT setup depicted on figure \ref{fig:HBT}a
would lead to the determination of single electron coherence in channel 1 from
current correlation measurements. 

\subsection{Quantum tomography: noise signals from single electron coherence}

In quantum optics, this idea has lead to the homodyne quantum tomography of the state of a single 
quantum electromagnetic mode: in this case, channel 2 is fed with a coherent monochromatic radiation called the local oscillator, whose phase 
is used as a control parameter \cite{Smithey:1993-1}. By analogy,  in the present situation, 
the ohmic contact on channel 2 will be used as a local oscillator since its chemical potential $\mu_{2}$ can 
be varied to scan the relevant energy range of single electron and hole excitations propagating along channel 1. 

\medskip

Contrary to the case of quantum optics experiments in the optical domain where the time resolved arrivals of single photons can
be observed, counting single electrons on sub-nanosecond time scales cannot be achieved today. Our protocol will
instead be based on 
the zero frequency component of the average over $\bar{t}$ of current correlations
\begin{equation}
\label{eq:tomography:measured}
S_{\alpha\beta}^{\mathrm{exp}}=2\int \overline{S_{\alpha,\beta}^{\mathrm{out}}(\bar{t}+\tau/2,\bar{t}-\tau/2)}^{\,\bar{t}}\,d\tau\,.
\end{equation} 
which are standard experimentally accessible quantities in quantum transport experiments. 
Equations \eqref{eq:QPC:S11} to \eqref{eq:QPC:S12} show
that these quantities now depends on the 
$\bar{t}$-average of the zero frequency component of the HBT contribution $\mathcal{Q}(t,t')$:
\begin{equation}
\label{eq:tomography:Q0}
\mathcal{Q}_{0}(\omega=0)=\int \overline{\mathcal{Q}(\bar{t}+\tau/2,\bar{t}-\tau/2)}^{\bar{t}}\,d\tau\,.
\end{equation}
Equation \eqref{eq:Q} shows that $\mathcal{Q}_{0}$ is nothing but the overlap between the single electron and hole
coherences of channels one and two. The idea of our tomography protocol is to find a suitable local oscillator
to be able to reconstruct $\mathcal{G}^{(e)}_{1}$ from measurements of this overlap.

Let us remark that $\mathcal{Q}_{0}(\omega=0)$ 
contains contributions associated with the two input ohmic contacts that do not depend on the source's contribution to single electron
coherence $\Delta\mathcal{G}^{(e)}_{1}$. Those are present even when the source is turned off and thus are not relevant for reconstructing $\Delta\mathcal{G}_{1}^{(e)}$.
The first one, given by $\mathcal{G}^{(e)}_{\mu_1}\mathcal{G}_{\mu_2}^{(h)}
+\mathcal{G}^{(h)}_{\mu_1}\mathcal{G}_{\mu_2}^{(e)}$ contributes to the partition noise $S^{\mathrm{exp}}_{1,\mathrm{part}}(\mu_{12})$ 
associated to the DC bias $\mu_{12}=\mu_{1}-\mu_{2}$ of the QPC \cite{Reznikov:1995-1,Saminadayar:1996-1}. 
The second one, given by $\mathcal{G}^{(e)}_{\mu_1}\Delta\mathcal{G}_{\mu_2}^{(h)}
+\mathcal{G}^{(h)}_{\mu_1}\Delta\mathcal{G}_{\mu_2}^{(e)}$ contributes to
the photoassisted noise $S_{1,\mathrm{pan}}^{\mathrm{exp}}[V_{\mathrm{ac}}(t)]$ due to the AC drive $V_{\mathrm{ac}}(t)$ theoretically predicted in 
\cite{Lesovik:1994-1,Pedersen:1998-2,Blanter:2000-1} and experimentally studied in 
\cite{Schoelkopf:1998-2, Reydellet:2003-1}. 
Because the single electron source generates no noise at zero frequency \cite{Mahe:2010-1}, 
the outcoming current noise in channel 1 can then be expressed in terms of the partition noise, 
the photoassisted noise and the excess HBT contribution we are looking for:
\begin{equation}
S_{11}^{\mathrm{exp}} = S^{\mathrm{exp}}_{11,\mathrm{part}}(\mu_{12})+
S_{11,\mathrm{pan}}^{\mathrm{exp}}[V_{\mathrm{ac}}(t)]
 + \mathcal{RT}\Delta\mathcal{Q}_{0}[\omega=0,\mu_{2},V_{\mathrm{ac}}(t)]\,.
\end{equation}
where $\Delta\mathcal{Q}_{0}[\omega=0,\mu_{2},V_{\mathrm{ac}}(t)]$ denotes the excess HBT contribution which depends on 
the source's contribution $\Delta\mathcal{G}^{(e)}_{1}$.
Thus, measuring the excess outcoming noise due to the source in one of the two channels directly gives access to the excess HBT contribution
which constitutes our experimental signal.  As we shall see now, it contains all the information needed to reconstruct the single electron coherence 
$\Delta\mathcal{G}^{(e)}_{1}$ emitted by the source.

\medskip

In the experimentally relevant case of a $T$ periodic source, $\Delta\mathcal{G}_{1}^{(e)}$ 
can be written as a Fourier transform with respect to $\tau=t-t'$ and 
a Fourier series with respect to $\bar{t}=(t+t')/2$. Therefore, single electron tomography
aims at reconstructing the harmonics $\Delta\mathcal{G}^{(e)}_{1,n}$ defined by:
\begin{equation}
\Delta\mathcal{G}_{1}^{(e)}(t,t')=\sum_{n=-\infty}^{+\infty}e^{-2\pi i n\bar{t}/T}\int \Delta\mathcal{G}^{(e)}_{1,n}(\omega)e^{-i\omega \tau}\frac{d\omega}{2\pi}\,.
\end{equation}
Let us first discuss the $n=0$ harmonic $v_{F}\Delta\mathcal{G}_{1,n=0}^{(e)}(\omega)$ which represents
the average density of electron excitations at energy $\hbar\omega$ emitted per period ($\Omega=0$ line on figure \ref{fig:coherence}b). 
As this quantity is an average over $\bar{t}$,
no homodyning is required and thus no AC voltage is applied: $V_{\mathrm{ac}}(t)=0$. 
At zero temperature, the variation of the experimental signal $\Delta\mathcal{Q}_{0}$
with $\mu_{2}$ reflects the single particle content of the source at the corresponding energy:
\begin{equation}
\label{eq:tomography:diagonal}
\frac{\partial (\Delta \mathcal{Q}_0)}{\partial \mu_2}[\omega=0,\mu_{2},V_{\mathrm{ac}}(t)=0]=-\frac{2e^2}{h}\,v_{F}\Delta\mathcal{G}_{1,n=0}^{(e)}(\mu_{2}/\hbar)\,.
\end{equation}
Indeed, if the potential $\mu_{2}$ becomes comparable to the energy of the emitted single electron state, the 
latter will always find an undistinguishable partner in the second incoming channel of the beam splitter so that the 
excess partition noise due to the source vanishes. This is reflected by the minus sign in the r.h.s of equation \eqref{eq:tomography:diagonal}.
Finally by varying the potential $\mu_2$, one can measure the energy distribution of single electron excitations in channel 1.

\medskip

Let us now consider the higher harmonics $\Delta\mathcal{G}_{1,n}^{(e)}(\omega)$ with $n\neq 0$ which contain the $\bar{t}$ dependence
of the single electron coherence. Accessing  $\Delta\mathcal{G}_{1,n}^{(e)}(\omega)$ requires homodyning 
the $\bar{t}$ dependence of $\Delta\mathcal{G}_{1}^{(e)}(t,t')$ at frequency $nf$ ($f=1/T$). This is achieved by applying an 
AC drive $V_{\mathrm{ac}}(t)=V\cos{(2\pi nf t+\phi)}$ to the Ohmic contact on channel 2. 
At zero temperature, the linear response $\bar{\chi}_{n}(\mu_{2},\phi)=[\partial (\Delta\mathcal{Q}_{0})/\partial(eV/nhf)]\vert_{\omega=0,V=0}$ 
to the AC drive of the excess HBT contribution of the source
is related to the single electron coherence by
\begin{equation}
\label{eq:tomography:non-diagonal}
\frac{\partial \bar{\chi}_n}{\partial \mu_2}(\mu_2,\phi)
=  \frac{e^2}{h}\,
\Re{\left[
e^{i\phi}\left(v_{F}\Delta\mathcal{G}^{(e)}_{1,n}(\frac{\mu_2}{\hbar}+\pi nf) 
 -  v_{F}\Delta\mathcal{G}^{(e)}_{1,n}(\frac{\mu_2}{\hbar}-\pi nf)\right)\right]}.
\end{equation}
Equations \eqref{eq:tomography:diagonal} and \eqref{eq:tomography:non-diagonal} 
form the central result of this paper: they relate the dependence of the experimental signals on the control parameters on
channel 2 (the chemical potential
$\mu_{2}$, the AC voltage amplitude $V$ and phase $\phi$) to the single particle coherence of the source. 
Inverting these relations basically leads to the reconstruction of the single electron coherence in frequency space and therefore
we call this procedure a {\em single electron quantum tomography protocol}. 

%\medskip
\subsection{Quantum tomography: proposed experimental procedure}
% GWENDAL

The experimental procedure reads as follows. First one measures the excess zero frequency partition noise $\Delta S_{11}^{\mathrm{exp}} = \mathcal{RT}\Delta\mathcal{Q}_{0}[\omega=0,\mu_{2},V_{\mathrm{ac}}(t)]$ by subtracting the zero frequency partition noise when the source is turned off. Then the $\mu_2$ dependence of $\Delta\mathcal{Q}_{0}$ is  measured by varying the chemical potential of the ohmic contact number 2.

To reconstruct the $n=0$ harmonic of the single electron coherence, no ac-drive is applied on ohmic contact 2. By numerical derivation of the $\mu_2$ dependence of the experimental data $\Delta\mathcal{Q}_{0}$, $\Delta\mathcal{G}_{1,n=0}^{(e)}(\mu_{2}/\hbar)$ is computed following equation \eqref{eq:tomography:diagonal}.

To reconstruct the $n \neq 0$ harmonics of the single electron coherence, the ac-drive  $V_{\mathrm{ac}}(t)=V\cos{(2\pi nf t+\phi)}$ is applied on ohmic contact 2. For $eV \ll n hf$, the measurement of  $\Delta\mathcal{Q}_{0}$ provides a direct determination of $\bar{\chi}_n$, as $\bar{\chi}_n(\mu_2,\phi) \approx \frac{\Delta\mathcal{Q}_{0}[\omega=0,\mu_{2},V_{\mathrm{ac}}(t)]}{ (eV/nhf)} $. By proceeding again to the numerical derivation of experimental data $\Delta\mathcal{Q}_{0}$, one gets the $\mu_2$ dependence of $\frac{\partial \bar{\chi}_n}{\partial \mu_2}(\mu_2,\phi)$.  It is computed for the two values $\phi=0$ and $\phi= \pi/2$ of the phase of the ac-drive, to provide information on the real and imaginary parts of $\Delta\mathcal{G}^{(e)}_{1,n}(\omega)$. Indeed, using equation \eqref{eq:tomography:non-diagonal}, one can relate adjacent values of the single electron coherence distant by $2 \pi nf$ :
\begin{eqnarray}
\Delta\mathcal{G}^{(e)}_{1,n}(\frac{\mu_2}{\hbar}) & = &   \frac{h}{e^2 v_{F}} \Big( \frac{\partial \bar{\chi}_n}{\partial \mu_2}(\mu_2 - \pi nf ,\phi=0) + i \, \frac{\partial \bar{\chi}_n}{\partial \mu_2}(\mu_2 - \pi nf,\phi=\pi/2) \Big)\nonumber\\
&  + &
\Delta\mathcal{G}^{(e)}_{1,n}(\frac{\mu_2}{\hbar}-2\pi nf)\,.
\end{eqnarray}
As $ \Delta\mathcal{G}^{(e)}_{1,n}(\mu_2 = \pm \infty) =0$, $ \Delta\mathcal{G}^{(e)}_{1,n}(\mu_2 ) $ can be measured step by step starting from  a point $\mu_2 = \mu_0$ where $\Delta\mathcal{G}^{(e)}_{1,n}(\mu_2=\mu_0)$ is known to vanish.

\medskip

To limit the total reconstruction time, an optimization strategy must be devised to choose the measurement points ({\it i.e.} the values of $\mu_{2}$)
so that regions where the coherence is expected to vary most are covered with maximal resolution. Such an optimization procedure is most conveniently 
performed when one has an idea of the expected experimental signal for the source to be characterized. This is why, in the next section, 
we will consider the problem of predicting the expected experimental signals.

\medskip

Before discussing signal predictions, let us consider temperature effects since in practice, the incoming channel has a finite 
electronic temperature $T_{\mathrm{el}}$. The corresponding formula are given in \ref{appendix:Temperature}
and show that the single electron coherence $\Delta\mathcal{G}_{1}^{(e)}$ can only be accessed with an energy resolution $k_{B}T_{\mathrm{el}}$. 
This stresses the necessity of working at the lowest possible temperature reachable by the experimental setup.

\section{Predictions for the single electron source}
\label{sec:electron-source}

To support the implementation of our single electron quantum tomography protocol,
we present predictions for the on demand single electron source demonstrated in \cite{Feve:2007-1}. 
In particular, we have computed the single electron coherence and the
experimentally accessible quantities $\Delta\mathcal{Q}_{0}(\mu_{2})=\Delta\mathcal{Q}_{0}[\omega=0,\mu_{2},V_{\mathrm{ac}}(t)=0]$ 
and $\bar{\chi}_{n}(\mu_{2},\phi)$. 

As suggested from previous studies of the average current \cite{Feve:2007-1} and of finite frequency current noise 
\cite{Mahe:2010-1,Moskalets:2008-1} of this source, in the experimentally relevant regime of operation, interaction effects within the dot can 
be neglected. The appropriate formalism to discuss free electron scattering in the presence of
a periodic drive is the Floquet theory \cite{Floquet:1883-1} which has been applied to quantum pumps
by Moskalets and B\"uttiker  \cite{Moskalets:2002-1,Moskalets:2007-1} and also 
to various driven nanoconductors by H¨\"anggi and his collaborators \cite{Kohler:2005-1}. More recently,
the fluctuations of the charge transferred by a mesoscopic turnstile have been predicted from Floquet theory by Battista and Samuelsson~\cite{Battista:2011-1}.

\subsection{Floquet approach to the mesoscopic capacitor}

Here we present the Floquet approach to the single electron coherence 
emitted by a driven single channel quantum conductor. Details specific
to the mesoscopic capacitor are given in \ref{appendix:Floquet}.

\medskip

\begin{figure}
\begin{center}
\includegraphics[width=9cm]{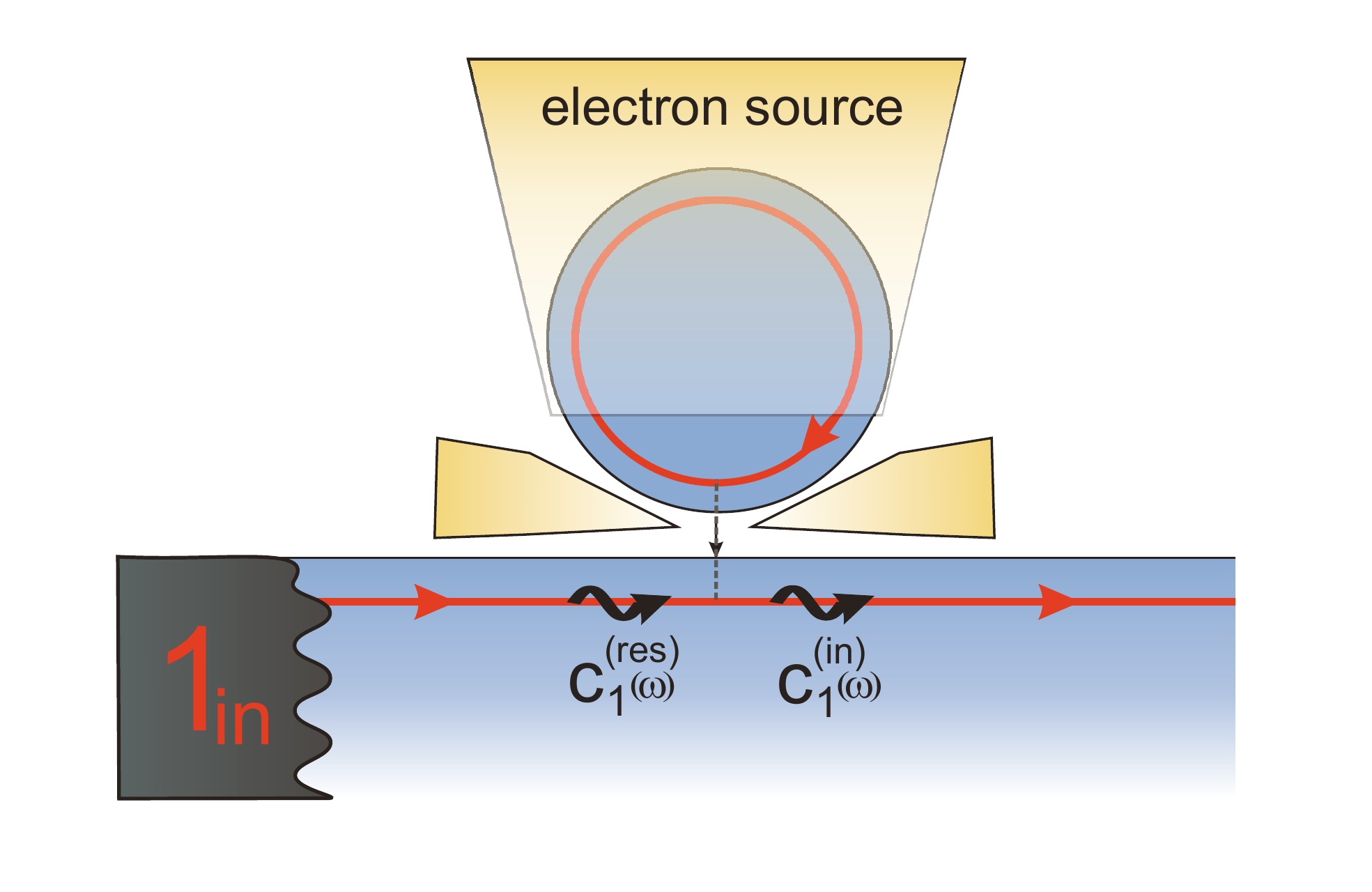}
\end{center}
\caption{\label{fig:Floquet} The incoming modes for the Floquet scattering formalism are the $c_{1}^{(\mathrm{res})}(\omega)$ 
issued by the reservoir (Ohmic contact) and we assume that the outcoming modes from the source are directly 
injected into the QPC of figure \ref{fig:HBT}a and thus are denoted by $c_{1}^{(\mathrm{in})}(\omega)$.}
\end{figure}

The Floquet scattering amplitude for electrons propagating through a driven quantum conductor is simply obtained as
\begin{equation}
\label{eq:Floquet:2}
\mathcal{S}_{\mathrm{Fl}}(t,t')=\exp{\left(\frac{ie}{\hbar}\int_{t'}^{t}V_{d}(\tau)\,d\tau\right)}\,\mathcal{S}_{0}(t-t')\,.
\end{equation}
where $V_{d}(\tau)$ is the periodic AC driving voltage applied to the dot and 
$S_{0}(t-t')$ is the scattering amplitude accross the undriven conductor, expressed in real time (see \ref{appendix:Floquet}). 
Knowing the Floquet scattering amplitude \eqref{eq:Floquet:2} leads to the real time 
single electron coherence emitted by the driven mesoscopic conductor:
\begin{equation}
\label{eq:Floquet:1}
\mathcal{G}^{(e)}_{1}(t,t')=\int \mathcal{S}_{\mathrm{Fl}}(t,\tau_{+})\,\mathcal{S}_{\mathrm{Fl}}(t',\tau_{-})^*\,\mathcal{G}_{\mu_{1}}^{(e)}(\tau_{+},\tau_{-})\,d\tau_{+}d\tau_{-}
\end{equation}
where $\mathcal{G}_{\mu_{1}}^{(e)}$ denotes the coherence function for electrons at chemical potential $\mu_{1}$. However as discussed before, 
we are interested in computing the single electron coherence in the frequency domain.  Therefore, we introduce the Floquet scattering matrix $\mathcal{S}_{\mathrm{Fl},n}(\omega)$
which represents the amplitude for photoassisted transitions between single electron states. It relates the single particle modes emitted from the reservoir 
$c^{(\mathrm{res})}_{1}(\omega)$ to the single electron modes emitted by the single electron source as shown
on figure \ref{fig:Floquet}a.
When the source is located close enough to the QPC, one expects decoherence and relaxation effects between the single electron source and the QPC 
of the setup of figure \ref{fig:HBT}a to be very weak. 
Assuming that they 
can be neglected, the modes emitted by the source can be identified with 
the incoming modes $c_{1}^{(\mathrm{in})}(\omega)$ of \eqref{eq:HBT:incoming}. We then have:
\begin{equation}
\label{eq:Floquet:scattering-definition}
c_1^{(\mathrm{in})}(\omega) = \sum_{n=-\infty}^{+\infty} \mathcal{S}_{\mathrm{Fl},n} (\omega) \,c_1^{(\mathrm{res})}(\omega + 2\pi n f)\,.
\end{equation}
Then, the $n$th harmonic $\mathcal{G}_{1,n}^{(e)}(\omega)$ can then be expressed under a form suitable for numerical computations:
\begin{equation}
\label{eq:Floquet:Gn}
v_{F}\mathcal{G}^{(e)}_{1,n} (\omega) = 
\sum_{k=-\infty}^{+\infty} \mathcal{S}_{\mathrm{Fl},k}\left(\omega+\pi nf\right)\mathcal{S}_{\mathrm{Fl},n+k}^*\left(\omega-\pi nf\right) 
\,n_F \left(\omega+(n+2k)\pi f\right)
\end{equation} 
where $n_{F}(\omega)$ is the Fermi distribution at chemical potential $\mu_{1}$ and temperature $T_{\mathrm{el}}$. 
Taking the Fourier transform of \eqref{eq:Floquet:2}, the Floquet scattering matrix $\mathcal{S}_{\mathrm{Fl},n}(\omega)$
can be computed in terms of the undriven conductor scattering matrix $\mathcal{S}_{0}(\omega)$ by:
\begin{equation}
\label{eq:Floquet:Floquet-matrix}
\mathcal{S}_{\mathrm{Fl},n}(\omega) = \sum_{k=-\infty}^{+\infty} C_k [V_{d}] C_{k+n}^*[V_{d}] \mathcal{S}_0 (\omega - 2\pi k f)
\end{equation} 
where the coefficients $C_{k}[V_{d}]$ denotes the Fourier transform of the phase accumulated by an electron experiencing the 
driving voltage $V_{d}(t)$ within the conductor:
\begin{equation}
\label{eq:Floquet:phase-harmonics}
\exp{\left( \frac{ie}{\hbar}\int_{-\infty}^t V_{d}(\tau) d\tau \right)} = \sum_{n=-\infty}^{+\infty} C_n [V_{d}] \,e^{-2\pi inf t}\,.
\end{equation}

\subsection{Numerical results}

\begin{figure*}
\includegraphics[angle=270,width=13cm]{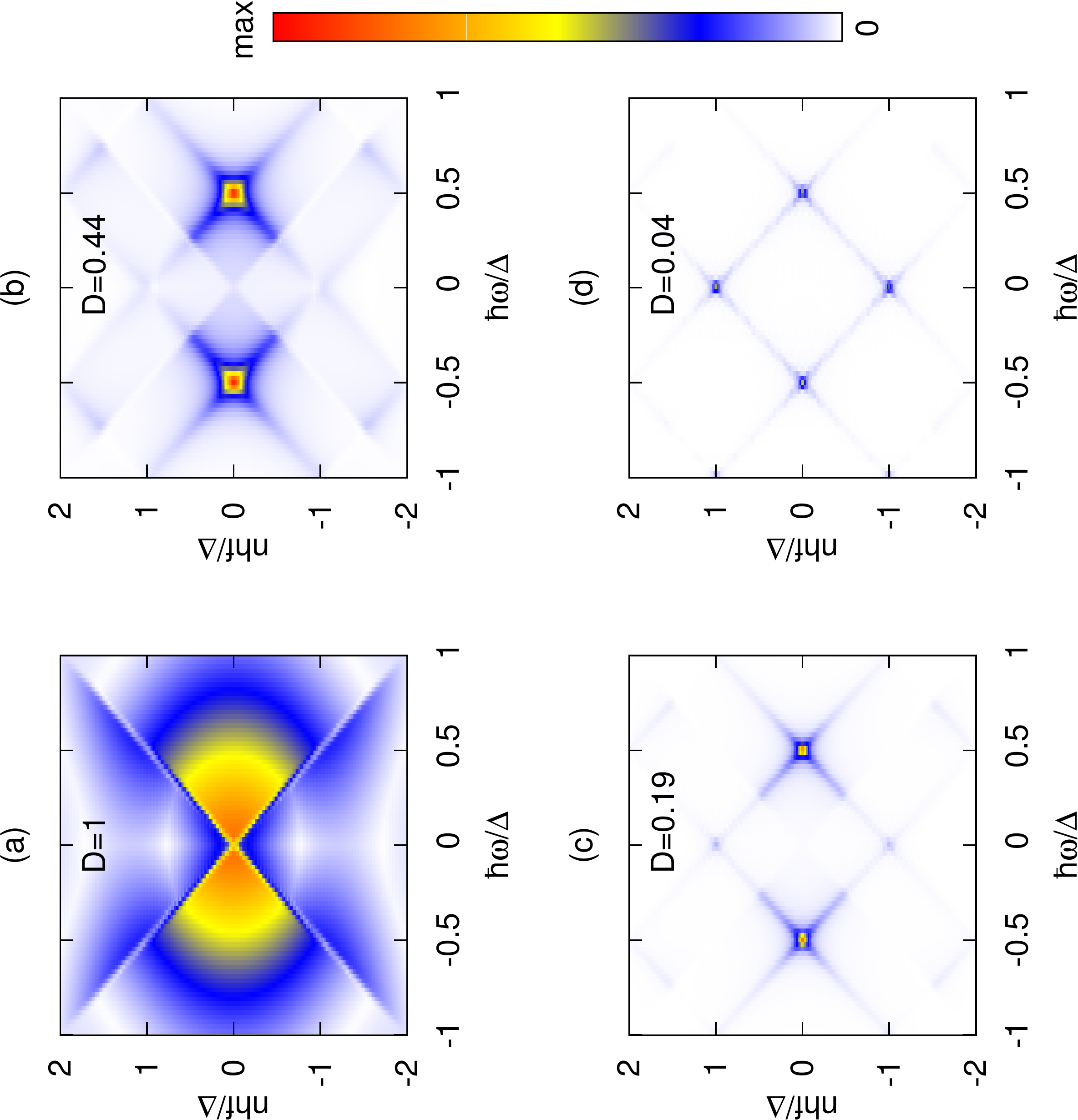}
\caption{\label{fig:tomo} 
Density plot visualization of $|v_{F}\Delta\mathcal{G}^{(e)}_{1,n}(\omega)|$ 
at $f=3~\mathrm{GHz}$, $T_{\mathrm{el}}\simeq 40\ \mathrm{mK}$ and $\Delta/k_{B}=4.7\ \mathrm{K}$ 
for dot to lead transmission (a) $D=1$, (b) $D=0.44$, (c) $D=0.19$ and (d) $D=0.04$.}
\end{figure*}

Figure \ref{fig:tomo} shows $|v_{F}\Delta\mathcal{G}^{(e)}_{1,n}(\omega)|$ plots 
for realistic values of the parameters of the mesoscopic capacitor: the level spacing is $\Delta/k_{B}=4.7~\mathrm{K}$, 
the electronic temperature is $T_{\mathrm{el}}\simeq 40~\mathrm{mK}$ and the driving frequency is $f=3~\mathrm{GHz}$. 
These results have been obtained by evaluating \eqref{eq:Floquet:Gn} numerically using a specific
form for $\mathcal{S}_{0}(\omega)$ already used to interpret the experimental data \cite{Gabelli:2006-1}, recalled in eq. \eqref{eq:FP} and
parametrized by the dot to lead transmission $D$. We have
considered the case of a square voltage and a number of tests have been performed on the numerical
results to ensure their validity (see \ref{appendix:Floquet}).

\medskip

\begin{figure*}
\includegraphics[angle=270,width=13cm]{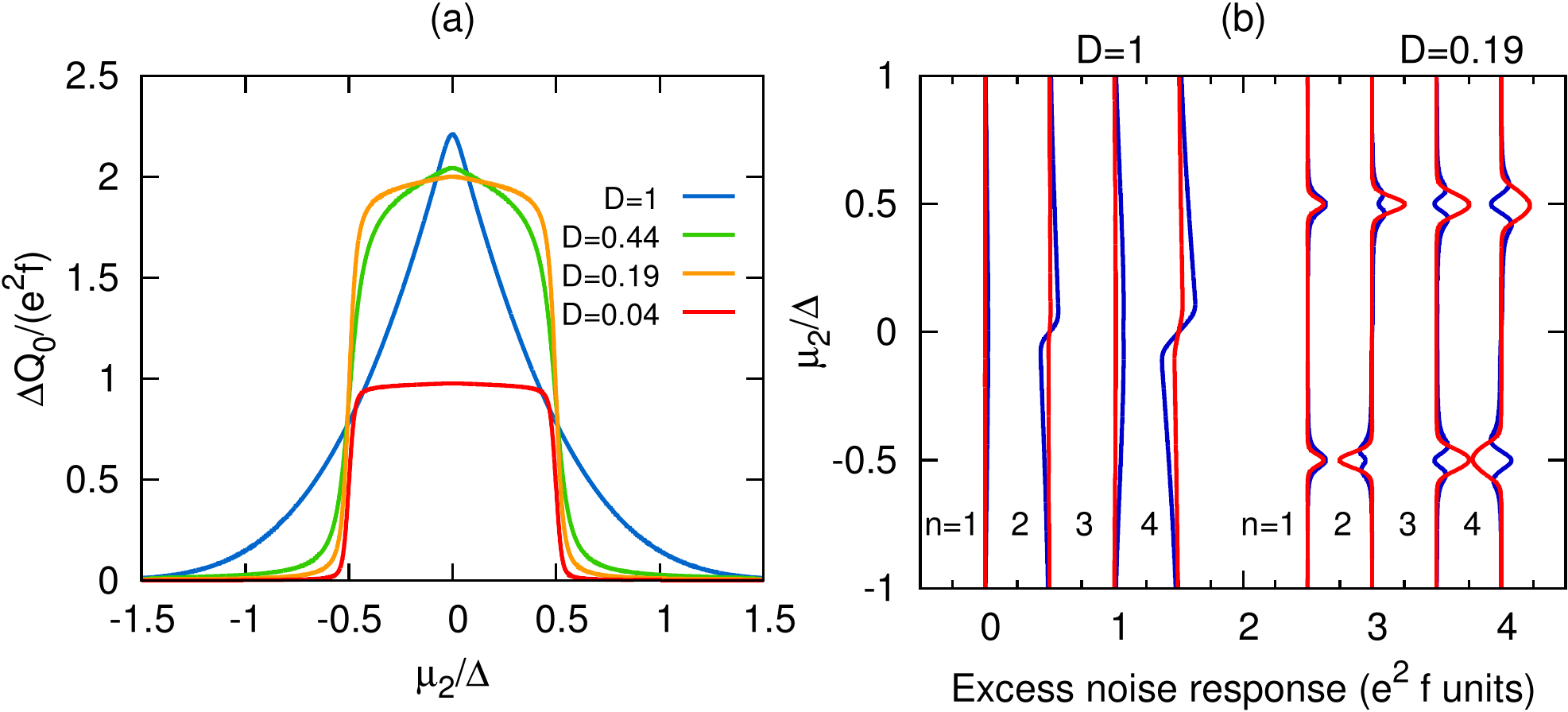}
\caption{\label{fig:tomo:signals} 
For the same values of $D$ as on figure \ref{fig:tomo}: (a) expected $\Delta\mathcal{Q}_{0}(\omega=0,\mu_{2})$ in function of
of $\mu_{2}/\Delta$ and in units of $e^2f\simeq 7.70\times10^{-29}\  \mathrm{A}^2\,\mathrm{Hz}^{-1}$; (b) Expected 
$\bar{\chi}_{n}(\mu_{2},\phi)$ for $n=1$ to $4$ with $\phi=0$ (blue curves) and $\phi=\pi/2$ (red curves) 
in function of $\mu_{2}/\Delta$ and in units of $e^2f$  for dot to lead transmission $D=1$ and $D=0.19$.}
\end{figure*}

When the dot is completely open ($D=1$), 
$\Delta\mathcal{G}^{(e)}_{1}$ 
presents strong electron/hole coherences and, within the electron and hole quadrants, is localized close to the
Fermi surface. The shape of
the experimental signal $\Delta\mathcal{Q}_{0}(\mu_{2})/e^2f$ depicted on figure \ref{fig:tomo:signals}a can then be
simply understood: an instantly responding system would lead to a triangular  $\Delta\mathcal{Q}_{0}(\mu_{2})$.
This is a direct consequence of the relation \eqref{eq:tomography:diagonal}
between the electron distribution function and $\Delta\mathcal{Q}_{0}(\mu_{2})$: in this situation, the square voltage would
shift a fraction of electrons of energies between $-\Delta/2$ and $0$ by $\Delta$, thus sending them above the Fermi surface and
giving rise to a triple step electron distribution function.
The smoothed shape of the scattering theory prediction is due to the finite temperature and to the finite frequency 
response of the edge channel at frequencies comparable to $h/\Delta$, the inverse of the time of flight around the dot.

\medskip

When $D$ decreases towards $0.19$, 
$\Delta\mathcal{G}_{1,n}^{(e)}(\omega)$ concentrates around the 
$n=0$, $\omega\simeq \pm \Delta/2\hbar$ points 
and simultaneously electron/hole coherences decrease, 
thus revealing energy resolved single electron and hole excitations. As we shall see in the next section, this is
where the mesoscopic capacitor behaves as a good single electron source.

\medskip

Pinching the dot even more ($D=0.04$) leads to a reappearance of 
electron/hole coherences (see Fig.~\ref{fig:tomo}d). 
In this regime, the source is driven too rapidly for single electron and hole excitations 
to fully escape the cavity in a half-period ($\gamma_{e} T\simeq 2.15$)  \cite{Feve:2007-1}. In fact, because at
the end of each half period, the electron or hole excitation to be emitted are still delocalized between the dot and the edge channel,
the source produces a linear combination of the many body states $|F_{\mu}\rangle$ and 
$\psi[\varphi_{h}]\psi^\dagger[\varphi_{e}]|F_{\mu}\rangle$ instead of a single electron hole pair state $\psi[\varphi_{h}]\psi^\dagger[\varphi_{e}]|F_{\mu}\rangle$. 
This is reflected by non vanishing electron hole coherences proportional 
to $\tilde{\varphi}_{e}(\omega_{+})\,\tilde{\varphi}_{h}(\omega_{-})^*$ corresponding to the spots in the (e/h) quadrants
of figure \ref{fig:tomo}d.

\medskip

Finally, in terms of current noise,  let us stress that the amplitude of the experimental signals depicted on figure \ref{fig:tomo:signals} 
is of the order of $10^{-29}\  \mathrm{A}^2\,\mathrm{Hz}^{-1}$, above the resolution of standard noise measurements. A resolution 
of a few  $10^{-30}\  \mathrm{A}^2\,\mathrm{Hz}^{-1}$ has already been obtained \cite{Mahe:2010-1,Parmentier:2010-1} in high 
frequency noise measurements using the electron emitter presented in this article. Noise floors below  
$10^{-30}\  \mathrm{A}^2\,\mathrm{Hz}^{-1}$ were even reported
in low frequency noise measurements of electron pumps \cite{Maire:2008-1} .

\subsection{Quality of the single electron source}

\begin{figure}[ht]
\includegraphics[width=12cm]{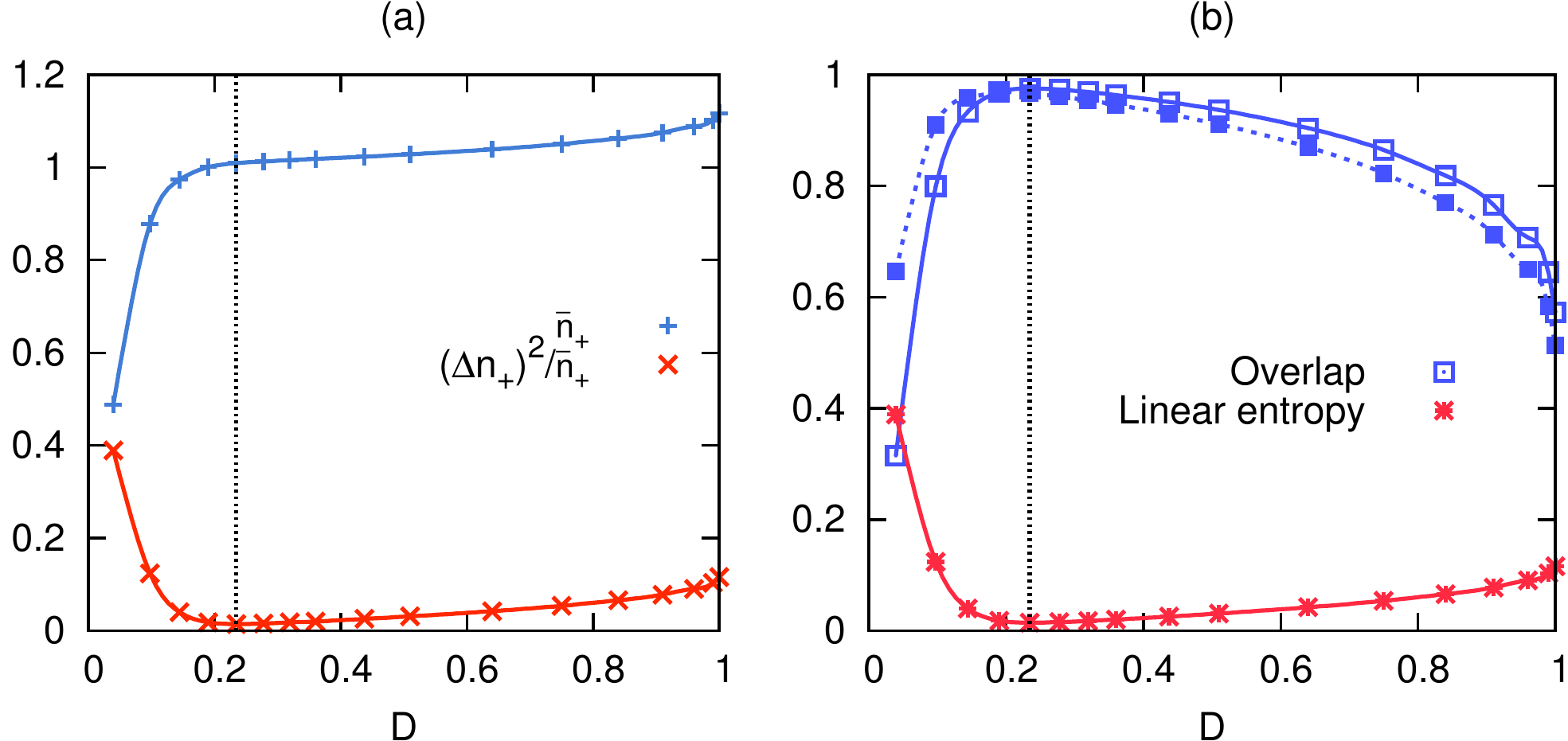}
\caption{\label{fig:quality} 
(a) Average number of electron excitations per cycle and its 
fluctuation as functions of $D$. (b) Linear entropy of electron excitations per emitted particle
and overlap of single electron coherence with the Lorentzian wavefunction as functions of $D$. 
The dashed curve gives the overlap per particle emitted. The optimal operating point corresponds to the vertical dotted line.
All curves are obtained for $f=3~\mathrm{GHz}$, $T_{\mathrm{el}}\simeq 40\ \mathrm{mK}$ and $\Delta/k_{B}=4.7\ \mathrm{K}$.}
\end{figure}

Measurements of single electron coherence of the source would lead to an assessment of its quality 
complementary to electron counting statistics \cite{Albert:2010-1}.
First, statistical properties of the source are described by the average number of electron excitations 
emitted per cycle and its fluctuation defined as the average value and fluctuation of the number of positive energy excitations:
\begin{equation}
\label{eq:quality:Ndef}
N_{+}=\int_{0}^{+\infty}c^\dagger(\omega)c(\omega)\,d\omega\,.
\end{equation}
where $c(\omega)$ and $c^\dagger(\omega)$ denote electron creation and destruction operators along the edge channel
fed by the source. The average value $\langle N_{+}\rangle$ is clearly given as an integral of the diagonal part 
of the single electron coherence of the source in the frequency domain. For the case of a periodic source
of period $T$ considered here, the average number $\overline{n}_+$ of electron excitations emitted per period is then given by:
\begin{equation}
\label{eq:quality:n+}
\bar{n}_+=T\int_{0}^{+\infty} v_{F}\Delta\mathcal{G}^{(e)}_{n=0}(\omega)\,\frac{d\omega}{2\pi}\,.
\end{equation}
Generically, the fluctuation $\langle N_{+}^2\rangle-\langle N_{+}\rangle^2$ involves a second order electronic coherence but
assuming that the source is described within Floquet scattering theory, Wick's theorem enables us to express the fluctuation of $N_{+}$
in terms of the single particle coherence. For a periodic source, the fluctuation $\Delta n_{+}$ of the number 
of electron excitations produced per period is then given as an
integral of single electron coherence over the (e/h) quadrants, thus stressing the role of coherent electron/hole pairs in fluctuations:
\begin{equation}
\label{eq:quality:delta-n+}
(\Delta n_{+})^2=T\sum_{n=1}^{+\infty}\int_{-\pi nf}^{\pi nf} \left| v_{F}\Delta\mathcal{G}_{n}^{(e)}(\omega)\right|^2\,\frac{d\omega}{2\pi}
\end{equation}
Scattering theory predictions for these quantities are depicted on figure~\ref{fig:quality}a as functions of the dot transparency.
When $D\rightarrow 1$, the electron number is not quantized: $\bar{n}_{+}$ is slightly greater than one and fluctuations are of the
order of $0.3$. In the shot noise regime where $D\ll 1$, the electron does not have the time to leave the quantum dot in time $T/2$ and
this translate into a decay of $\bar{n}_{+}$ and an increase of relative fluctuations $(\Delta n_{+})^2/\bar{n}_{+}\rightarrow 1/2$ consistent with
predictions from the probabilistic model of \cite{Albert:2010-1}. In the intermediate region, quantization of the emitted number of electron
excitations per period is observed:  at $D\simeq 0.22$, we find that $\bar{n}_{+}\simeq 1.009$ and $(\Delta n_{+})^{2}/\bar{n}_{+}\simeq 0.014$ ($\Delta n_{+}\simeq 0.12$). 

This is the quantum jitter regime where the current noise
reflects the randomness of electron emission through quantum tunneling \cite{Mahe:2010-1}. In this regime, almost certainly one electron and one hole
are emitted during each period \cite{Albert:2010-1}.
From a statistical point of view, the optimal point for single electron emission is when $\bar{n}_{+}\simeq 1$ and $\Delta n_{+}$ is minimal
which occurs for $D\simeq 0.22$ with our choice of parameters.

Next, to assess the source's quantum coherence, we propose to use  the linear entropy of the reduced
density operator for electron excitations as well as 
its overlap of $\Delta\mathcal{G}^{(e)}_{1}$ with the Lorentzian wave function expected 
in a discrete level model. The linear entropy 
measures how far the reduced density operator for electron excitations 
departs from a pure state \cite{Book:Nielsen-Chuang}. The overlap with a given electron wave function gives the
probability that, per cycle, an electron be detected in this single particle state. For a periodic source, the linear entropy
production per emitted electron is given by:
\begin{equation}
\label{eq:quality:entropy}
S_{L}=1-\frac{T}{\bar{n}_{+}}
\sum_{n=-\infty}^{+\infty}
\int_{|n|\pi f}^{+\infty} |v_{F}\Delta\mathcal{G}^{(e)}_{n}(\omega)|^2\,\frac{d\omega}{2\pi}\,.
\end{equation}
In the same way the fidelity with respect
to a normalized electron wave function $\varphi_{e}$ expressed  in the frequency domain as 
$\tilde\varphi_{e}(\omega)=\int \varphi_{e}(x)e^{i\omega x/v_F}\,dx$ can be
computed in terms of the single electron coherence:
\begin{equation}
\label{eq:quality:fidelity}
\mathcal{F}(\mathcal{G}^{(e)}|\varphi_{e})
=\sum_{n=-\infty}^{+\infty}
\int_{|n|\Omega_{T}}^{+\infty}
\Delta\mathcal{G}^{(e)}_{n}(\omega)\,\tilde\varphi_{e}^*(\omega+n\pi f)\tilde\varphi_{e}(\omega-n\pi f)\,
\frac{d\omega}{2\pi}\,.
\end{equation}
In the case of the single electron source, it is natural to choose as a trial wave function $\tilde\varphi_e(\omega)$ a truncated Lorentzian 
wavefunction \eqref{eq:lorentzian} representing the result of the decay from a resonant level at energy $\hbar\omega_{e}=\Delta/2$ into 
the semi infinite continuum of accessible electron states. 

\medskip

Predictions for the linear entropy \eqref{eq:quality:entropy} and for the overlap 
\eqref{eq:quality:fidelity} with this resonant level wavefunction are depicted on
figure \ref{fig:quality}b for experimentally reasonable parameters. In this case, we see that
the best operating point is obtained for $D\simeq 0.22$. At this optimal point, the source is predicted to be highly coherent and 
well described by the discrete level model wavefunctions. In particular, the fidelity per emitted electron between the reduced density 
operator for electron like excitations and the resonant level wavefunction is $0.97$ and the purity is $0.99$.
As stated before, decreasing $D$ does not leave enough time for 
emitting single electron and hole excitations. This  leads to the generation of electron/hole coherence
responsible for quantum fluctuations of $N_{+}$ and also for lower purity of single electron and hole
excitations.

\section{Conclusion}
\label{sec:conclusion}

To conclude, we have proposed a quantum tomography protocol to reconstruct the quantum state of single electron
excitations in quantum Hall edge channels. Its implementation would
provide a complete assessment of the quantum coherence of single electron sources, either energy resolved \cite{Feve:2007-1,Leicht:2011-1,Battista:2011-1} or
time resolved \cite{Keeling:2006-1}. In particular, by
probing harmonics $\Delta\mathcal{G}_{n}^{(e)}$ up to $nf=10\ \mathrm{GHz}$ or more, it would give access to the 
individual electronic wavepackets on a sub nanosecond time scale. 

This experimental breakthrough could 
then be used for quantitative studies of decoherence and relaxation of single electron excitations \cite{Degio:2009-1} complementary to
recent studies of non equilibrium electronic relaxation in quantum Hall edge channels \cite{LeSueur:2010-1,Degio:2010-1}. A new generation 
of experiments aiming at the controlled manipulation of the
quantum state of single to few electrons could then be envisioned in the near future. In particular, new experiments could
involve decoherence engineering as in \cite{Roulleau:2009-1} where a voltage probe increases decoherence at will 
within a Mach-Zehnder interferometer. An important issue is quantum coherence protection as in  \cite{Altimiras:2010-2} where
electronic decoherence is limited through an appropriate sample design. Another exciting although challenging 
perspective is to combine electron coherence measurements, photon statistics measurement \cite{Zakka:2010-1} and 
single electron sources in order to explore the non classicality of photons radiated by an electric current carried by trains 
of coherent electrons \cite{Beenakker:2004-1}. 

\begin{acknowledgements}
The authors thank M. B\"uttiker, Ch. Glattli, T. Jonckheere, F. Pierre, J.M. Raimond and J. Rech for useful discussions, references and remarks.
This work is supported by the ANR grant ''1shot'' (ANR-2010-BLANC-0412).
\end{acknowledgements}

\medskip

\appendix
\section{Finite temperature effects on single electron tomography}
\label{appendix:Temperature}

Let us consider the case of a source at finite electronic temperature $T_{\mathrm{el}}$ on channel 2. Then, equation 
\eqref{eq:tomography:diagonal} becomes
\begin{equation}
\label{eq:HBT:tomography:occupation-number:nonzero-temp}
\left(\frac{\partial \Delta \mathcal{Q}_{0}}{\partial \mu_2}\right)(\mu_{2},T_{\mathrm{el}})
 =  -\frac{2e^2}{h}
\int_{-\infty}^{+\infty} \frac{v_{F}\Delta\mathcal{G}_{1,n=0}^{(e)}\left(\frac{\mu_{2}}{\hbar}+\frac{k_BT_{\mathrm{el}}}{\hbar}\,x\right)}{4\cosh^2{(x/2)}}\,dx
\end{equation}
whereas, equation \eqref{eq:tomography:non-diagonal} has to be replaced by:
\begin{equation}
\label{eq:HBT:tomography:finite-T}
\left(\frac{\partial \bar{\chi}_n}{\partial \mu_2}\right)(\mu_{2},T_{\mathrm{el}}) = \frac{e^2}{h}\,\int_{-\infty}^{+\infty}
\frac{G(n,\phi,\mu_{2}+k_{B}T_{\mathrm{el}}x) }{4\cosh^2{(x/2)}}\,dx,
\end{equation}
where
\begin{equation}
G(n,\phi,\mu) = \Re{\left[e^{i\phi}\left(v_{F}\mathcal{G}^{(e)}_{1,n}(\frac{\mu}{\hbar}+\pi nf)-
v_{F}\mathcal{G}^{(e)}_{1,n}(\frac{\mu}{\hbar}-\pi nf)\right)\right]} \,.
\end{equation}
As is clear for these expressions, the single tomography protocol would then reconstruct a convolution of the single electron coherence in channel 1 with
a thermal smearing function of width $k_{B}T_{\mathrm{el}}$. 

\section{The driven mesoscopic capacitor}
\label{appendix:Floquet}

The mesoscopic capacitor is modeled as in ref.~\cite{Gabelli:2006-1} as an electronic Fabry-Perot cavity whose transparency is modulated by 
the transmission $D$ of its QPC. 
Here $\Delta$ denotes the level spacing within the quantum dot (electronic cavity). 
The scattering matrix for the undriven mesoscopic capacitor is
given by \cite{Gabelli:2006-1}:
\begin{equation}
\label{eq:FP}
\mathcal{S}_{0}(\varepsilon)=\frac{\sqrt{1-D}-e^{2\pi i\varepsilon /\Delta}}{1-\sqrt{1-D}\,e^{2\pi i\varepsilon/\Delta}}\,.
\end{equation}
This choice assumes that in the absence of drive, the Fermi level of channel $1$
is equidistant from two consecutive energy levels of the dot, a situation that can always be realized by applying an appropriate DC voltage. 
Here the electron escape time from the dot is given by $\gamma_{e}=(\Delta/h)\times 2D/(2-D)$ \cite{Mahe:2010-1}.

In principle, our formalism can be applied for any drive voltage. In particular, we could implement the precise form
generated by the voltage generator used in the experiment taking into account its limitations. But in the present paper
we have considered that a $T$ periodic square drive voltage is applied to the mesoscopic capacitor:
$V(t)=V_{d}$ for $0\leq t<T/2$ and $V(t)=-V_{d}$ for $-T/2\leq t<0$ and we have chosen its amplitude $2eV_{d}$ to be equal to the level spacing $\Delta$.
In this case, the Fourier coefficients of the driving phase are then given by:
\begin{eqnarray}
\label{eq:Floquet:Cn}
C_n [V] & = &
\frac{2}{\pi} \sin{\left( \frac{\pi}{2} (\mathcal{A}-n) \right)}
\frac{\mathcal{A}}{\mathcal{A}^2-n^2} e^{i(n\pi/2 - \mathcal{A})} 
\ \mathrm{for}\ n\ \neq \mathcal{A}, \\
 & = & 1/2\ \mathrm{otherwise} \ ,
\end{eqnarray}
with $\mathcal{A}= eV_{d}/2hf$. 

\medskip

The images produced for this paper represent $|\Delta\mathcal{G}^{(e)}_{n}(\omega)|$ 
for $|n|\leq 100$ and $|\hbar\omega|\leq 2\Delta$. Numerical convergence was achieved by summing over 4000 harmonics. Various tests
have been performed such as hermiticity $\mathcal{G}_n^{(e)} (\omega)^*= 
\mathcal{G}_{-n}^{(e)} (\omega)$ and electron hole symmetry
$\Delta\mathcal{G}^{(e)}_{n}(-\omega)=(-1)^{n+1}\Delta\mathcal{G}^{(e)}_{n}(\omega)$. 
The total charge emitted per cycle has been computed and
neutrality has been checked up to $10^{-5}$. 
Finally, the Cauchy-Schwarz inequality satisfied by $\mathcal{G}^{(e)}$
has also been checked numerically as well as the behaviour with respect of time translation of the driving signal.

\end{document}